\documentclass[preprint,12pt,authoryear]{elsarticle}

\usepackage{amssymb}
\usepackage{url}
\usepackage{subfig}

\journal{JPDC}

\usepackage{pgfplots}
\usepackage{tikz}
\usetikzlibrary{arrows,shadows,backgrounds,calc,decorations.pathreplacing,decorations.pathmorphing}

\newcommand\QTfigure[6]   
{
	\draw
	($ (0 cm,0.5cm-.5*#2 cm) + (0 cm,-#5*.1 cm) +(#3cm,-#4cm)$) node[QT,minimum width=#2cm,draw]
	(Q#1) {$#1$}
	;
} 
\newcommand\QTCfigure[6]   
{
	\draw
	($ (1.7 cm,.5cm-.5*#2 cm) + (0 cm,-#5*.1 cm) +(#3cm,-#4cm)$) node[QTC,minimum width=#2cm,minimum height=5cm,draw]
	(Q#1) {$#1$}
	;
}

\begin{document}

\begin{frontmatter}


\title{A figure of merit for describing\\the performance of scaling of parallelization}

\author[label1]{J\'anos V\'egh}
\ead{J.Vegh@uni-miskolc.hu}
\author[label2]{P\'eter Moln\'ar}%
\ead{pmolnar@lib.unideb.hu}
\author[label1]{J\'ozsef V\'as\'arhelyi}
\ead{vajo@mazsola.iit.uni-miskolc.hu}

\address[label1]{University of Miskolc, Hungary}
\address[label2]{PhD School of Informatics, University of Debrecen, Hungary}

\begin{abstract}
With the  spread of multi- and many-core processors
more and more typical task is to re-implement some source code written originally for a single processor
to run on more than one cores.
Since it is a serious investment, it is important to decide how much efforts pays off,
and whether the resulting implementation has as good performability as it could be.
The Amdahl's law  provides some theoretical upper limits for the performance gain
reachable through parallelizing the code, but it needs the detailed architectural knowledge
of the program code, does not consider the housekeeping activity needed for parallelization
 and cannot tell how the actual stage of parallelization implementation performs.
The present paper suggests a quantitative measure for that goal. This figure of merit is derived experimentally,
from measured running times, and number of threads/cores.
It can be used to quantify the used parallelization technology, 
the connection between the computing units, the acceleration technology under the
given conditions, or the performance of the software team/compiler.

\end{abstract}


\end{frontmatter}


\section{Introduction}\label{sec:introduction}

\noindent 
The computer manufacturing technology is not any more able to produce quicker processors, see \cite{ClockVsIPC2000}.
The crisis of the computing experienced since cca. year 2000, see \cite{ComputingPerformance:2011},
increased the demand towards parallel computing.

On hardware side: "\emph{Processor and network architectures are making rapid progress with more and more cores being integrated into single processors and more and more machines getting connected with increasing bandwidth. Processors become heterogeneous and reconfigurable.}", see \cite{SoOS:2010}.
On software side: "\emph{parallel programs ... are notoriously difficult to write, test, analyze, debug, and verify, much more so than the sequential versions}", see \cite{ReliableParallel2014}. 
In addition, the typical real-life programs show complex parallelization behavior, see  \cite{YavitsMulticoreAmdahl2014},
and also the apparently massively parallel algorithms can behave 
extremely ineffectively, see \cite{Pingali:2011:TaoOfParallelism}.
Different merits are derived for characterizing parallel systems, from 
simple speedup to price per performance~\cite{Karp:parallelperformance1990}.

Because all of these difficulties, demands and possibilities, uncountable developments are running and planned to re-implement some existing code, written having a single processor in mind, for the already ubiquitos multi-core processors. To find out whether it is worth to invest in such efforts for parallelization, as well as to decide where to stop the development, one needs some quantitative measure of the parallelism. Even today, in the "multicore era", see  \cite{HillMulticoreAmdahl2008}, the performance is described by 
(a modified version of) Amdahl's law, see \cite{AmdahlSingleProcessor67}. 
Unfortunately, Amdahl's law provides
no support for the goals mentioned: it needs information on code architecture, makes assumptions which are not any more valid for modern accelerated processors and applications heavily using operating system services.
However, scrutinizing the conditions and reverting the Amdahl's formula, it is possible to derive such a figure of merit.

%
\def\scalefact{0.67}
\tikzstyle{QT} = [top color=white, bottom color=blue!30,
minimum height={0.8},rotate=90,scale=\scalefact,xshift=-.5cm,yshift=-.5cm,
draw=blue!50!black!100, drop shadow] 
\tikzstyle{QTC} = [top color=green!80, bottom color=white,
minimum height={0.8},rotate=90,scale=\scalefact,xshift=-.5cm,yshift=-.8cm,
draw=blue!50!black!100, drop shadow] 
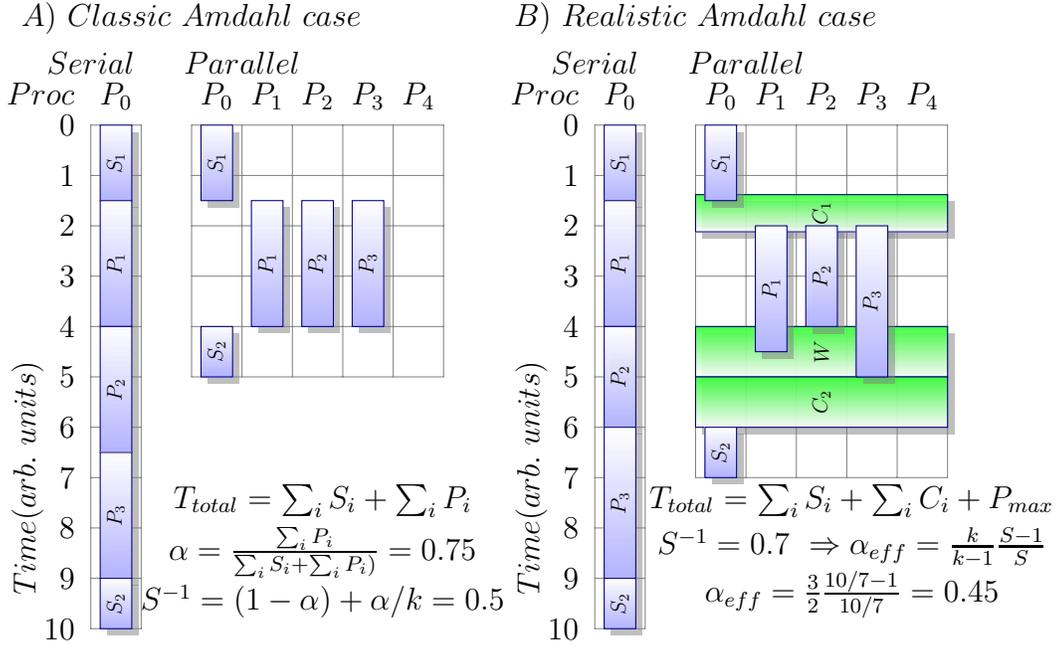
\begin{figure*} 
	\begin{tikzpicture}[scale=\scalefact,cap=round]
	
	\draw[style=help lines,step=1] (0,-10) grid (1,0);
	\node[right,above] at (2cm,1.6cm) {$A)\ Classic\ Amdahl\ case$};
	\node[right,above] at (0cm,.8cm) {$Serial$};
	\node[right,above] at (-1cm,.2cm) {$Proc$};
	\node[rotate=90,above=.5] at (-.7cm,-7cm) {$Time (arb.\ units)$};
	
	\foreach \x/\xtext in {0,...,0}
	\draw[xshift=\x cm+.5 cm,yshift=0.1cm]  node[above=.1]
	{$P_{\xtext}$};
	
	\foreach \y/\ytext in {0,...,10}
	\draw[yshift=-\y cm,xshift=-0.1cm]  node[left]
	{${\ytext}$};
	
	\QTfigure{S_1}{1.5}{0}{0}{0}{0}  %
	\QTfigure{P_1}{2.5}{0}{0}{15}{0}  %
	\QTfigure{P_2}{2.5}{0}{0}{40}{0}  %
	\QTfigure{P_3}{2.5}{0}{0}{65}{0}  %
	\QTfigure{S_2}{1}{0}{0}{90}{0}  %
	
	\draw[style=help lines,step=1] (2,-5) grid (7,0);
	\node[right,above] at (3cm,.8cm) {$Parallel$};
	\foreach \x/\xtext in {0,...,4}
	\draw[xshift=\x cm+2.5 cm,yshift=0.1cm]  node[above=.1]
	{$P_{\xtext}$};
	%
	\QTfigure{S_1}{1.5}{2}{0}{0}{0}  %
	\QTfigure{P_1}{2.5}{3}{0}{15}{0}  %
	\QTfigure{P_2}{2.5}{4}{0}{15}{0}  %
	\QTfigure{P_3}{2.5}{5}{0}{15}{0}  %
	\QTfigure{S_2}{1}{2}{0}{40}{0}  %
	
	\node[right,above] at (4.6cm,-8cm) {$T_{total}=\sum_i S_i + \sum_i P_i$};
	\node[right,above] at (4.6cm,-9.2cm) {$\alpha=\frac{\sum_iP_i} {\sum_i S_i + \sum_i P_i)} =0.75$};
	\node[right,above] at (4.6cm,-10cm) {$S^{-1}=(1-\alpha) +\alpha/k = 0.5$};
	
	\draw[style=help lines,step=1] (10,-10) grid (11,0);
	\node[right,above] at (12cm,1.6cm) {$B)\ Realistic\ Amdahl\ case$};
	\node[right,above] at (10cm,.8cm) {$Serial$};
	\node[right,above] at (9cm,.2cm) {$Proc$};
	\node[rotate=90,above=.5] at (9.3cm,-7cm) {$Time (arb.\ units)$};
	
	\foreach \x/\xtext in {0,...,0}
	\draw[xshift=\x cm+10.5 cm,yshift=0.1cm]  node[above=.1]
	{$P_{\xtext}$};
	
	\foreach \y/\ytext in {0,...,10}
	\draw[yshift=-\y cm,xshift=-0.1cm+10cm]  node[left]
	{${\ytext}$};
	\QTfigure{S_1}{1.5}{10}{0}{0}{0}  %
	\QTfigure{P_1}{2.5}{10}{0}{15}{0}  %
	\QTfigure{P_2}{2.0}{10}{0}{40}{0}  %
	\QTfigure{P_3}{3.0}{10}{0}{60}{0}  %
	\QTfigure{S_2}{1}{10}{0}{90}{0}  %
	
	\draw[style=help lines,step=1] (12,-7) grid (17,0);
	\node[right,above] at (13cm,.8cm) {$Parallel$};
	\foreach \x/\xtext in {0,...,4}
	\draw[xshift=\x cm+12.5 cm,yshift=0.1cm]  node[above=.1]
	{$P_{\xtext}$};
	%
	\QTCfigure{C_1}{.5}{12}{0}{15}{0}  %
	\QTCfigure{W}{1}{12}{0}{40}{0}  %
	\QTCfigure{C_2}{1}{12}{0}{50}{0}  %
	
	\QTfigure{S_1}{1.5}{12}{0}{0}{0}  %
	\QTfigure{P_1}{2.5}{13}{0}{20}{0}  %
	\QTfigure{P_2}{2.0}{14}{0}{20}{0}  %
	\QTfigure{P_3}{3.0}{15}{0}{20}{0}  %
	\QTfigure{S_2}{1}{12}{0}{60}{0}  %

	\node[right,above] at (15.1cm,-8cm) {$T_{total}=\sum_i S_i + \sum_i C_i + P_{max}$};
	\node[right,above] at (15.1cm,-9cm) {$S^{-1}= 0.7\  \Rightarrow\alpha_{eff} = \frac{k}{k-1}\frac{S-1}{S} $};
	\node[right,above] at (15.1cm,-10cm) {$\alpha_{eff} =  \frac{3}{2}\frac{10/7 - 1}{10/7} = 0.45$};

	\end{tikzpicture}
	\caption{Illustrating Amdahl's law for idealistic and realistic cases}
	\label{fig:amdahl}
\end{figure*}

\section{\uppercase{Limitations of parallelization}}


\subsection{Amdahl's law}\label{sec:amdahlslaw}

Amdahl's considerations focus on the fact that some parts ($P_i$) of a code can be parallelized,
some ($S_i$) must remain sequential. He also mentioned that data housekeeping
causes some overhead, which in his paper was estimated to be in the range 20\% to 40\%,
and  that \emph{the nature of this overhead appears to be sequential}.

Although Amdahl just wanted to draw the attention to the limitations
of the single-processor approach applied to large-scale computing, his followers
also provided a widely used formula, focussing on the parallelizable part of the code.
As the left side of Fig~\ref{fig:amdahl} demonstrates, the usual interpretation implies three essential
restrictions:
\begin{itemize}\setlength\itemsep{0em}
	\item the parallelized parts are of equal length in terms of execution time
	\item the housekeeping (controling the parallelization, passing parameters,
	exchanging messages, etc.) has no costs in terms of execution time
	\item the number of parallelizable chunks coincides with the number of available computing resources
\end{itemize}
\noindent Essentially,  this is why \emph{Amdahl's law represents a theoretical upper limit for parallelization gain}.
In Fig~\ref{fig:amdahl} the left side shows the idealistic case where the original process in the single-processor system comprises the sequential only parts $S_i$, and the parallelizable parts $P_i$. 
One can also see that the control components $C_i$ are of the same nature as $S_i$,
the non-parallelizable components. This also means that even in the idealistic case
when $S_i$ are negligible, $C_i$ will represent a bound for parallelization.
From the figure the meaning of $\alpha$ is: in what fragment of time (in terms of the 
time needed for completely sequential processing)
the helper processors are utilized. When the task is scaled to several processors, the goal is 
obviously to maximize the utilization of the helper processors.

The realistic case (shown in the right side of Fig~\ref{fig:amdahl}) however, is, that
the parallelized parts are \emph{not} of equal length (even if they contain exactly the same 
instructions, the hardware operation in modern processors may execute them in 
considerably different times (for examples see the operation of hardware accelerators inside a core or the network operation between processors, etc.)
and also that the time required to control parallelization is not negligible and varying.
Here $\alpha_{eff}$ provides a value for an \emph{average utilization} of the helper cores.
Obviously, the unused cores and the unbalanced load of cores degrades
this average utilization. To characterize the effects like 
sharing the processing between different number of helper cores
(\emph{the performance of the scaling}) or using different hardware conditions, one needs a quantitative 
figure of merit.

The figure also calls the attention to the fact that the static correspondence between program chunks and
processing units can be very inefficient: all assigned processing units must wait for the delayed unit
and also the capacity is lost if the number of computing resources exceeds the number of the parallelized chunks.

\subsection{Factors affecting parallelism}

Usually,  Amdahl's law is expressed with the formula 
\begin{equation}
S^{-1}=(1-\alpha) +\alpha/k
\end{equation}

\noindent where $k$ is the number of parallelized code fragments, 
$\alpha$ is the ratio of the parallelizable part  to the total sequential part,
$S$ is the measurable speedup. 
The assumption can be visualized that (assuming many processing units)
in $\alpha$ fraction of the running time the processors are processing,
in (1-$\alpha$) fraction they are waiting. I.e. $\alpha$ describes
how much, in average, the processors are utilized.
Having those data, the resulting speedup can be estimated.

For a system under test, where  $\alpha$ is not \textit{a priory} known,
one can derive from the measurable speedup  $S$ 
an \emph{effective parallelization} factor as

\begin{equation}
\alpha_{eff} = \frac{k}{k-1}\frac{S-1}{S}\label{equ:alphaeff}
\end{equation}

\noindent where $S$ is now the measured speedup, and  $k$ 
is the number of the available cores.
Obviously,  for the classical case, $\alpha = \alpha_{eff}$; which simple means that
in \emph{idealistic} case the actually measurable effective parallelization 
reaches the theoretically possible one.
In other words, $\alpha$ describes a system the \emph{architecture} of which is completely known,
$\alpha_{eff}$ describes a system the \emph{performance} of which is known from experiments.
Again in other words,  $\alpha$ is the \emph{theoretical upper limit}, which can hardly be reached,
while $\alpha_{eff}$ is the \emph{experimental actual value}, which describes the complex architecture and the actual conditions. It is interesting to note, that $\alpha_{eff}$ is an 
absolute measure of utilizing the available processing capacity, see section \ref{sec:practical}.
Numerically ($1-\alpha_{eff}$) equals with the $f$ value, established theoretically by~\cite{Karp:parallelperformance1990}.

The $\alpha_{eff}$ can then be used to refer back to the Amdahl's classical
assumption even in the realistic case when the parallelized chunks 
have different length and the overhead to organize parallelization is not negligible.
Note that in case of real tasks a kind of Sequential/Parallel Execution Model, see \cite{YavitsMulticoreAmdahl2014}, shall be applied, which cannot use 
the simple picture reflected by $\alpha$, but $\alpha_{eff}$ gives a good merit
of the degree of parallelization for the duration of  the execution of the process,
and can be compared to the results of the technology-dependent parametrized formulas.

With our notations, in the classical Amdahl case on the left side in Fig.~\ref{fig:amdahl}

\begin{equation}
S= \frac{\sum_i S_i  + \sum_i P_i}{\sum_i S_i + \max_i P_i} = 2
\end{equation}

\noindent and 

\begin{equation}
\alpha = \alpha_{eff}= \frac{\sum_i Pi}{\sum_i S_i+\sum_i Pi} = 3/4
\end{equation}

Now we can compare the effective parallelization in the two cases shown in Fig.~\ref{fig:amdahl}.
In the realistical case $S =10/7$, which results in \begin{equation}
\alpha_{eff}= \frac{3}{2}\frac{10/7 - 1}{10/7} = 0.45
\end{equation}

\begin{figure}
	\begin{tikzpicture}  [scale=1.1]
	\begin{axis}[
	ylabel=Sequential ratio ,
	xlabel=Overhead ratio, 
	zlabel=$\alpha_{eff}$,
	zmax=1,
	grid=both,
	colormap/jet,
	]
	\addplot3[
	mesh, 
	samples=11,
	domain=0:0.2,y domain=0:0.6,
	]
	{ 
		(3./(3-1))* 
		( 1.- ((x+.25*(1.+y))/(x+3.*.25)))
	};
	\addplot3[
	mesh, scatter,
	samples=11,
	domain=0.2:0.4,y domain=0:0.6,
	]
	{ 
		(3./(3-1))* 
		( 1.- ((x+.25*(1.+y))/(x+3.*.25)))
	};
	\addplot3[
	mesh, 
	samples=11,
	domain=0.4:0.8,y domain=0:0.6,
	]
	{ 
		(3./(3-1))* 
		( 1.- ((x+0.25*(1.+y))/(x+3.*0.25)))
	};
	\addlegendentry{$\alpha_{eff}\ k=3, \max_i P_i=0.25$}
	
	\addplot3+[only marks] coordinates {
		(0.6,0.25, 0.45) };
	\end{axis}
	\end{tikzpicture}
	\caption{Behavior of the effective parallelization $\alpha_{eff}$ in function of the overhead ratio (relative to the parallelizable payload execution length) and the ratio of the sequential part (relative to the total sequential execution time).}
	\label{EffectiveFactor}
\end{figure}
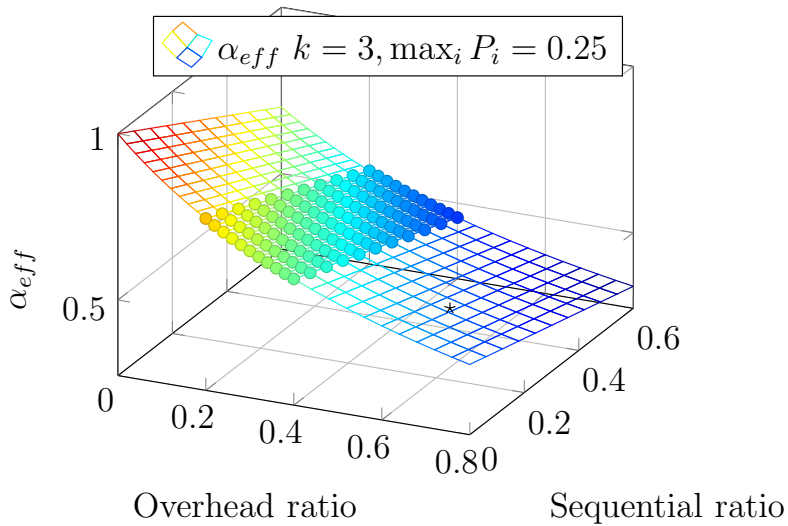

\noindent As seen, the overhead and the different duration of the parallelized parts
reduced the effective parallelization drastically relative to the theoretically reachable value. 
Fig~\ref{EffectiveFactor} gives a feeling on the effect of the computer system behaviour
on the effective parallelization.  The middle region (marked by balls) is mentioned by Amdahl as typical range of overhead. The asterisk  in the figure shows the "working point"
corresponding to the values used in Fig~\ref{fig:amdahl}. 

One can see that the effective parallelization drops quickly with both increasing 
overhead and sequential parts of the program. This fact draws the attention to the idea that \emph{through 
	decreasing either the control time or the sequential-only fraction of the code (or both), and utilizing
	the otherwise wasted processing capacity, a serious gain in the effective parallelization
	can be reached}. This was experienced in the "dynamic" architecture
by \cite{HillMulticoreAmdahl2008}.

\subsection{Different implementations of parallelism}

The timing analysis in Fig \ref{fig:amdahl} can be applied to different kinds of parallelizations,
from the processor-level parallelization (instruction or data level parallelization, in the nanoseconds range)
to OS-level parallelization (including thread-level parallelization using several processors or cores, in the microseconds range),
to network-level  (between networked computers, like grids, in the milliseconds range).
The principles are the same, see \cite{SynchronizationEverything2013},
independently of the kind of implementation.
In agreement with \cite{YavitsMulticoreAmdahl2014},
housekeeping overhead is always present (and mainly depends on the architectural solution), and remains a key question,
although the mains focus is always on reducing its effect.

The actual speedup (or the effective parallelization) depends strongly on the 'tricks' used during implementation.
Although HW and SW parallelism are interpreted differently, they even can be combined, see
\cite{ChandyParallelism1999}, resulting in hybrid architectures. 
For those greatly different architectural solutions it is even more hard to interpret  $\alpha$,
while $\alpha_{eff}$ allows to compare
different implementations (or the same implementation under different conditions) in such cases, too.

Notice that in all kinds of parallelization the relative overhead fulfills the observation made by Amdahl:
for better performability the overhead time cannot exceed dozens of percents relative to the execution time of the parallelized chunk.
Between networked computers the control times $C_i$ 
are in the millisecond range and a lot of data must be transmitted, so for making parallelization efficient, typically complete jobs
(having duration sometimes even in the minutes range)
are sent to the other computers. When using thread-level parallelizm through OS services inside the computer,
the "expenses" of organizing threads is in the order of thousands of instructions,
and typically the length of the working threads is also at least in that order or above. 
In hardware level parallelization, when using hyperthreading, the values of $C_i$ are in the range of 1 clock cycle.
However, the program chunks $P_k$ are also in the order of a few clock cycles,
i.e. in a comparable range. Similar holds for the speculative evaluation,
the out-of-order evaluation, etc. 

\section{Practical applications}\label{sec:practical}

The good metric to select describing parallelism depends on many factors, see~\cite{Karp:parallelperformance1990}. 
The newly introduced metric $\alpha_{eff}$ describes how effectively the computing task
is distributed between the processing units.
As outlined above, the  control functionality as well as unequalities in
cutting program into equally long pieces (including data transfer between processing units)
degrade $\alpha_{eff}$.
Since $1-\alpha$ gives the sequential-only part of the program, 
$(1-\alpha_{eff})$ is expected to describe the ratio of the total
(even unintended) sequential part, i.e. it is a sensitive measure of
disturbances of parallelization. 
Since a larger load imbalance results in a larger decrease in value of $\alpha_{eff}$
(as can be concluded from~\cite{Karp:parallelperformance1990}, 
$\alpha_{eff}$ is a kind of derivative of relative speedup),
problems can be identified better than from speedup values.
If the parallelization is well-organized (load balanced, small overhead, right number of processors), $\alpha_{eff}$  saturates at unity, so tendencies can be better displayed
through using $(1-\alpha_{eff})$.

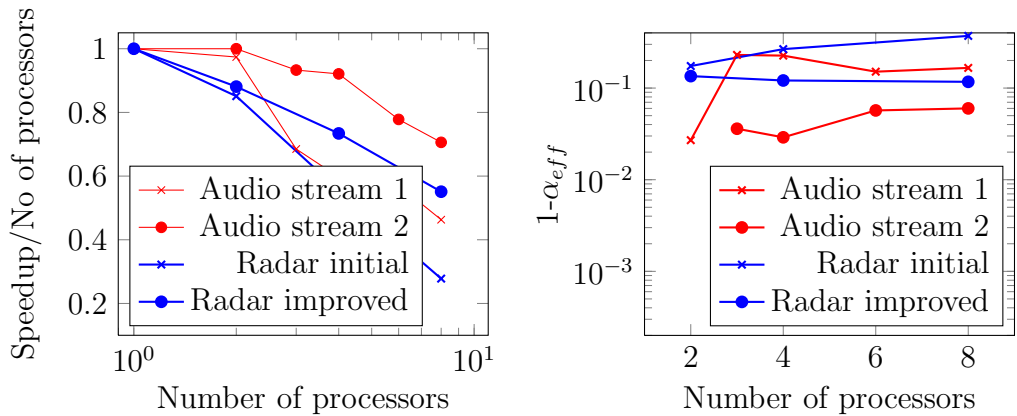
\begin{figure}[]
	\pgfplotsset{width=6.5cm}
	\centering
	\subfloat
	{
		\begin{tikzpicture}
		\begin{axis}[
		legend style={
			cells={anchor=east},
			legend pos={south west},
		},
		xmin=.9, xmax=11,
		ymin=0.1, ymax=1.05, 
		xlabel=Number of processors,
		ylabel=Speedup/No of processors,
		xmode=log,
		]
		\addplot[ mark=x,red] plot coordinates {
			(1, 1) 
			(2, .974)    
			(3, .685) 
			(4, .597)
			(6, .571) 
			(8, .463) 
		};
		\addlegendentry{Audio stream 1}
		\addplot[ mark=*,red] plot coordinates {
			(1, 1) 
			(2, 1)    
			(3, .933) 
			(4, .921)
			(6, .778) 
			(8, .706) 
		};
		\addlegendentry{Audio stream 2}
		
		\addplot[ mark=x,blue,thick] plot coordinates {
			(1, 1) 
			(2, .851)    
			(4, .556)
			(8, .278) 
		};
		\addlegendentry{Radar initial}
		\addplot[ mark=*,blue,thick] plot coordinates {
			(1, 1) 
			(2, .881)    
			(4, .734)
			(8, .551) 
		};
		\addlegendentry{Radar improved}

		\end{axis}
		\end{tikzpicture}		
} \subfloat
{
		\begin{tikzpicture}
		\begin{axis}[
		legend style={
			cells={anchor=east},
			legend pos=south east,
		},
		xmin=1, xmax=9,
		ymin=.0002, ymax=0.4, 
		xlabel=Number of processors,
		ylabel={1-$\alpha_{eff}$},
		ymode=log,
		]				
		
		\addplot[ thick, color=red,mark=x]
		plot coordinates {
			(2, 0.027) 
			(3, .23) 
			(4, .226) 
			(6, .151) 
			(8, .166) 
		};
		\addlegendentry{Audio stream 1}

		\addplot[ thick, color=red,mark=*]
		plot coordinates {
			(2, 0) 
			(3, .036) 
			(4, .029) 
			(6, .057) 
			(8, .060) 
		};
		\addlegendentry{Audio stream 2}
		
		\addplot[ thick, color=blue,mark=x]
		plot coordinates {
			(2, 0.174) 
			(4, .266) 
			(8, .371) 
		};
		\addlegendentry{Radar initial}
		\addplot[ thick, color=blue,mark=*]
		plot coordinates {
			(2, 0.135) 
			(4, .121) 
			(8, .117) 
		};
		\addlegendentry{Radar improved}

		\end{axis}
		\end{tikzpicture}
	}
	\caption{Relative speedup (left side)
	and ($1-\alpha_{eff}$) (right side) values, measured running the audio and radar processing on
	 different number of cores. \protect{\cite{CompilerInfrastructure:2014}}
		\label{fig:CompilerDiagram}}
\end{figure}

\subsection{Characterizing parallelization efforts}
In paper~\cite{CompilerInfrastructure:2014} a compiler making effective parallelization 
of an originally sequential code for different number of cores
is described and validated by running the executable code on platforms having the corresponding number of cores.
Let us apply Equ. (\ref{equ:alphaeff}) to their results, shown 
in Figs 8 and 10 in their paper.

Fig. \ref{fig:CompilerDiagram} left side displays efficiency
(Efficiency = speedup divided by the number of cores) in function of 
number of cores for two different processings of audio streams, and 
for two processings of radar signals.
The data displayed in the figures are derived simply through reading back
diagram values from the mentioned figures in~\cite{CompilerInfrastructure:2014},
so they may be not accurate. However, they are accurate enough to support our conclusions.

Based on their merit, the authors of \cite{CompilerInfrastructure:2014} can only declare
a qualitative statement, that the 'efficiency'  decreases less steadily (dots on the figure) with the growing
number of cores "The higher number of parallel processes in Audio-2 gives better results"), if they consider load balancing.

It can surely be stated, that the improvement was successful:
in both cases the decrease with increasing number of cores is less steep.
The diagrams cannot tell, however, whether further improvements are possible
or whether the parallelization is uniform in function of the number of cores.
In contrast, the $(1-\alpha_{eff})$ diagrams (right side) show also, that
in both cases the improvement decreased the sequential part, i.e. improved
the parallelization. It can also be seen, that in the case of audio stream,
the parallelization is improved and so did the uniformity of parallelization.
In the case of radar signals, without optimization the parallelization decreases
as the number of the cores increases. With load balancing option on,
the parallelization is at any core number gets better.
The compiler really does a good job: $\alpha_{eff}$ is practically constant,
the compiler finds nearly all possibilities:


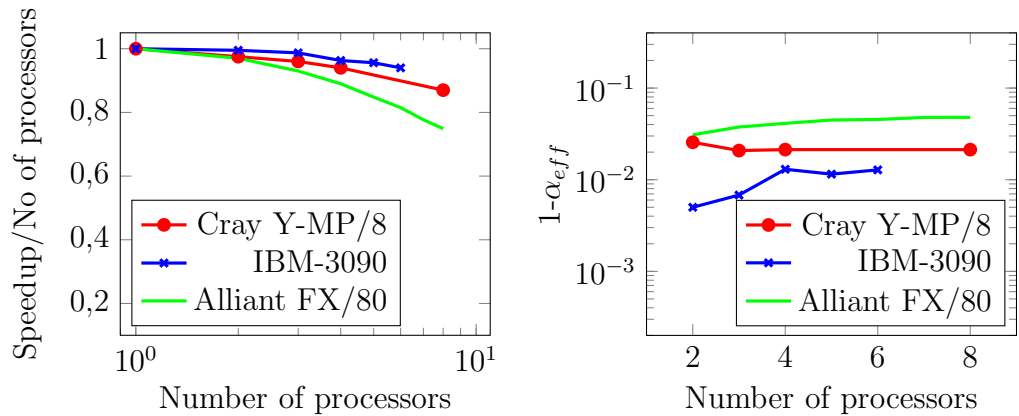
\begin{figure}[]
	\pgfplotsset{width=6.5cm}
	\centering
	\subfloat
	{
		\begin{tikzpicture}
		\begin{axis}[
		/pgf/number format/.cd,
		use comma,
		1000 sep={},
		legend style={
			cells={anchor=east},
			legend pos=south west,
		},
		xmin=.9, xmax=11,
		ymin=0.1, ymax=1.05, 
		xlabel=Number of processors,
		ylabel=Speedup/No of processors,
		xmode=log,
		]
		\addplot[ very thick, color=red,mark=*] plot coordinates {
			(1,1) 
			(2,0.975)  
			(3,0.960) 
			(4, 0.940)
			(8, 0.870) 
		};
		\addlegendentry{Cray Y-MP/8}
		\addplot[ very thick, color=blue,mark=x] plot coordinates {
			(1,1) 
			(2, 0.995) 
			(3, 0.987)
			(4, 0.963)
			(5, 0.956)
			(6, 0.940)
		};
		\addlegendentry{IBM-3090}
		\addplot[ very thick, color=green,mark=y] plot coordinates {
			(1,1) 
			(2,0.970)  
			(3, 0.930) 
			(4, 0.890)
			(5, 0.848) 
			(6, 0.815) 
			(7, 0.777)
			(8, 0.749) 
		};
		\addlegendentry{Alliant FX/80}
		
		\end{axis}
		\end{tikzpicture}		
} \subfloat
{
		\begin{tikzpicture}
		\begin{axis}[
		/pgf/number format/.cd,
		use comma,
		1000 sep={},
		legend style={
			cells={anchor=east},
			legend pos=south east,
		},
		xmin=1, xmax=9,
		ymin=.0002, ymax=0.4, 
		xlabel=Number of processors,
		ylabel={1-$\alpha_{eff}$},
		ymode=log,
		]
		\addplot[ very thick, color=red,mark=*] plot coordinates {
			(2,0.0256)  
			(3,0.0208) 
			(4,0.0213)
			(8,0.0213) 
		};
		\addlegendentry{Cray Y-MP/8}
		\addplot[ very thick, color=blue,mark=x] plot coordinates {
			(2, 0.0050) 
			(3, 0.0068)
			(4, 0.0130)
			(5, 0.0115)
			(6, 0.0128)
		};
		\addlegendentry{IBM-3090}
		\addplot[ very thick, color=green,mark=y] plot coordinates {
			(2 ,0.0309)  
			(3, 0.0376) 
			(4, 0.0412)
			(5, 0.0448) 
			(6, 0.0454) 
			(7, 0.0478)
			(8, 0.0479) 
		};
		\addlegendentry{Alliant FX/80}
		
		\end{axis}
		\end{tikzpicture}
	}

	\caption{Relative speedup (left side)
	and ($1-\alpha_{eff}$) (right side) values, measured running Linpack on different computers with different number of parallel
	processors. \protect{\cite{Karp:parallelperformance1990}}
		\label{fig:KarpDiagram1}}
\end{figure}

Note that the absolute values in the  two cases must not be compared:
they represent the sequenctial-only part of the two programs,
and they might be different for the different programs.
The uniformity of the values make also highly probable, that in the case of audio
streams further optimization can be done, at least for the 2-core and 3-core systems,
while processing of radar signals reached its bounds.
In addition, it can also be estimated, that the non-parallelizable part amounts to $\approx < 10\%$.

Notice that using $S$ and $\alpha_{eff}$ are simply two different points of view of the same thing.
If we have the information, how big is the $\alpha$ fraction of the code which can be executed in parallel
(an architectural point of view), we can estimate the maximum speedup we can reach. Here we assume that all processors have
the same architecture.
The experimentalist's point of view is different: if we can measure the speedup,
and know how many processing units was used, we can estimate how big $\alpha_{eff}$ 
fraction was running in parallel (assuming the mentioned ideal conditions).
Notice also that in deriving $\alpha_{eff}$ no assumption was made on the code architecture or the nature  
of the computing units or their way of linking, so the merit can be used to characterize the effect of the \emph{change}, if one of the mentioned components changes. It means, one can use that merit
for describing the effect of changing the code architecture, or the (behavior of the)  interconnecting network,
the (internal architecture of the) hardware setup, etc. as well.

\subsection{Characterizing HW architectures}

In Table I of~\cite{Karp:parallelperformance1990}, different architectures are compared,
running the same program (Linpack) on computers from different manufacturers and having
different number of processors. Because the subject of the paper was deriving a metric
from measured data, here the precision of the values is much better. The high degree
of parallelization results in $\alpha_{eff}$ values, close to unity, so the value 
 ($1-\alpha_{eff}$) is used in Figure~\ref{fig:KarpDiagram1}.
 
As also in the previous case, the efficiency decreases with the increasing number of cores.
The effective parallelization is nearly constant, and the difference in the absolute values
can be attributed to implementation details of the different computers.

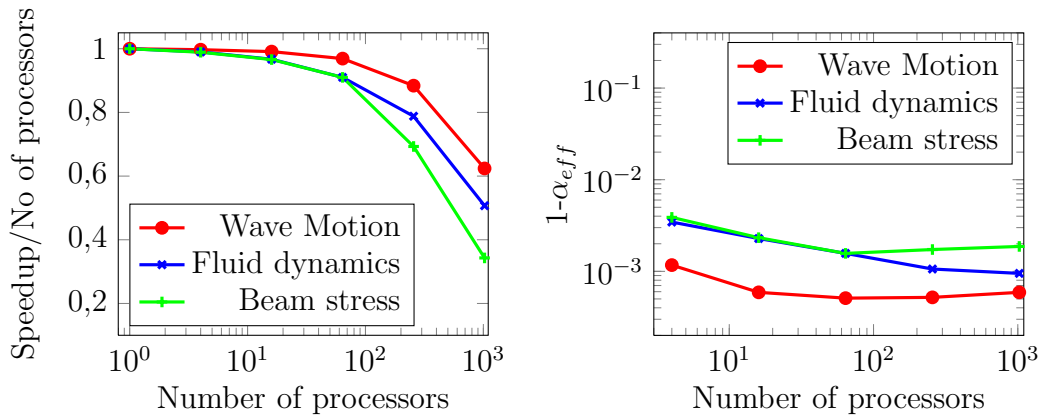
\begin{figure}[]
	\pgfplotsset{width=6.5cm}
	\centering
	\subfloat
	{
		\begin{tikzpicture}
		\begin{axis}[
		/pgf/number format/.cd,
		use comma,
		1000 sep={},
		legend style={
			cells={anchor=east},
			legend pos=south west,
		},
		xmin=0.8, xmax=1100,
		ymin=0.1, ymax=1.05, 
		xlabel=Number of processors,
		ylabel=Speedup/No of processors,
		xmode=log,
		]
		\addplot[ very thick, color=red,mark=*] plot coordinates {
			(1,1) 
			(4,0.997)  
			(16, 0.991) 
			(64, 0.969)
			(256, 0.884) 
			(1024, 0.624) 
		};
		\addlegendentry{Wave Motion}
		\addplot[ very thick, color=blue,mark=x] plot coordinates {
			(1,1) 
			(4,0.990)  
			(16, 0.967) 
			(64, 0.910)
			(256, 0.788) 
			(1024, 0.507) 
		};
		\addlegendentry{Fluid dynamics}
		\addplot[ very thick, color=green,mark=+] plot coordinates {
			(1,1) 
			(4,0.989)  
			(16,0.966) 
			(64, 0.910)
			(256, 0.693) 
			(1024, 0.343) 
		};
		\addlegendentry{Beam stress}
		
		\end{axis}
		\end{tikzpicture}		
} \subfloat
{
		\begin{tikzpicture}
		\begin{axis}[
		/pgf/number format/.cd,
		use comma,
		1000 sep={},
		legend style={
			cells={anchor=east},
			legend pos=north east,
		},
		xmin=3, xmax=1100,
		ymin=.0002, ymax=0.4, 
		xlabel=Number of processors,
		ylabel={1-$\alpha_{eff}$},
		xmode=log,
		ymode=log,
		]
		\addplot[ very thick, color=red,mark=*] plot coordinates {
			(4, 0.00117)  
			(16, 0.00059) 
			(64, 0.00051)
			(256, 0.00052) 
			(1024, 0.00059) 
		};
		\addlegendentry{Wave Motion}
		\addplot[ very thick, color=blue,mark=x] plot coordinates {
			(4, 0.00345)  
			(16, 0.00228) 
			(64, 0.00157)
			(256, 0.00106) 
			(1024, 0.00095) 
		};
		\addlegendentry{Fluid dynamics}
		\addplot[ very thick, color=green,mark=+] plot coordinates {
			(4, 0.00388)  
			(16, 0.00233) 
			(64, 0.00157)
			(256, 0.00173) 
			(1024, 0.00187) 
		};
		\addlegendentry{Beam stress}
		
		\end{axis}
		\end{tikzpicture}
	}

	\caption{Relative speedup (left side)
	and ($1-\alpha_{eff}$) (right side) values, measured running different algorithms
	on the same computer with different number of parallel processors. \protect{\cite{Karp:parallelperformance1990}}}
		\label{fig:KarpDiagram2}
\end{figure}

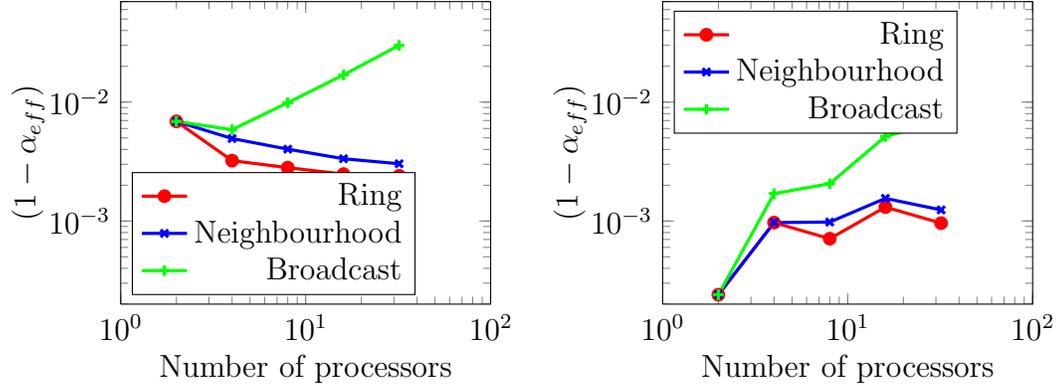
\begin{figure}[]
	\pgfplotsset{width=6.5cm}

	\subfloat
	{
		\begin{tikzpicture}
		\begin{axis}[
		/pgf/number format/.cd,
		use comma,
		1000 sep={},
		legend style={
			cells={anchor=east},
			legend pos=south west,
		},
		xmin=1, xmax=100, 
		ymin=0.0002, ymax=0.07,
		xlabel=Number of processors,
		ylabel=$(1-\alpha_{eff})$,
		xmode=log,
		ymode=log,
		]
		\addplot[ very thick, color=red,mark=*] plot coordinates {
			(2,0.00688)  
			(4, 0.00322) 
			(8, 0.00281)
			(16, 0.00248) 
			(32, 0.00240) 
		};
		\addlegendentry{Ring}
		\addplot[ very thick, color=blue,mark=x] plot coordinates {
			(2,0.00688)  
			(4, 0.00494) 
			(8, 0.00402)
			(16, 0.00334) 
			(32, 0.00303) 
		};
		\addlegendentry{Neighbourhood}
		\addplot[ very thick, color=green,mark=+] plot coordinates {
			(2,0.00688)  
			(4, 0.00586) 
			(8, 0.00988)
			(16, 0.01692) 
			(32, 0.03002) 
		};
		\addlegendentry{Broadcast}
		
		\end{axis}
		\end{tikzpicture}		
} \subfloat
{
		\begin{tikzpicture}
		\begin{axis}[
		/pgf/number format/.cd,
		use comma,
		1000 sep={},
		legend style={
			cells={anchor=east},
			legend pos=north west,
		},
		xmin=1, xmax=100, 
		ymin=0.0002, ymax=0.07,
		xlabel=Number of processors,
		ylabel=$(1-\alpha_{eff})$,
		xmode=log,
		ymode=log,
		]
		\addplot[ very thick, color=red,mark=*] plot coordinates {
			(2,0.00024)  
			(4, 0.00097) 
			(8, 0.00071)
			(16, 0.00131) 
			(32, 0.00096) 
		};
		\addlegendentry{Ring}
		\addplot[ very thick, color=blue,mark=x] plot coordinates {
			(2,0.00024)  
			(4, 0.00097) 
			(8, 0.00098)
			(16, 0.00155) 
			(32, 0.00124) 
		};
		\addlegendentry{Neighbourhood}
		\addplot[ very thick, color=green,mark=+] plot coordinates {
			(2,0.00024)  
			(4, 0.00170) 
			(8, 0.00206)
			(16, 0.00514) 
			(32, 0.00682) 
		};
		\addlegendentry{Broadcast}
		\end{axis}
		\end{tikzpicture}
	}

	\caption{($1-\alpha_{eff}$) values, measured when minimizing Rosenbrock function (left side) and 
	Rastrigin function (right side),
	on the same SoC, using different communication strategies, in function of the used processors. \protect{\cite{ReconfigurableAdaptive2016}}}
		\label{fig:SoCCommunicationDiagram}
\end{figure}

\subsection{Comparing communication strategies in Systems-on-Chip}

In their work~\cite{ReconfigurableAdaptive2016} the authors compare
different communication stategies  their PSO uses when minimizing the 
Rosenbrock function and the Rastrigin function, respectively.
As it could be expected, in the case of the 'broadcast' type communication
the 'sequential' fraction increases with the number of the processors,
in other cases practically remains constant. The fluctuation
shows the limitations of the (otherwise excellent) measurement precision.

It is worth to compare Fig.~\ref{fig:SoCCommunicationDiagram} with
Fig~\ref{fig:KarpDiagram2}, right side. All the three diagrams 
show the scaling behavior of some procedures, in function of the 
processing units. It would be worth to run the processes shown in 
 Fig~\ref{fig:KarpDiagram2}, in the PSO, to find out the advantage
 of having the processing units \textit{inside} the chip.

\subsection{Characterizing scaling of parallelization}
In Table II of~\cite{Karp:parallelperformance1990}, execution time of different programs 
are given in function of processors. The data are shown in Fig.~\ref{fig:KarpDiagram2}. 
As presented there, the efficiency drops in a catastrophic way as the number of cores
increases, while ($1-\alpha_{eff}$) changes only within the limits of the 
measurement error. Notice that Figs. \ref{fig:CompilerDiagram}, \ref{fig:KarpDiagram1} and \ref{fig:KarpDiagram2}
use the same scale, and that the steeper decrease of efficiency means
higher values of ($1-\alpha_{eff}$).

The behavior of efficiency deserves some analyzis. As detailed at the beginning of section \ref{sec:amdahlslaw}, 
the distinguished constituent in Amdahl's classic analysis is the parallelizable fraction $\alpha$,
all the rest (including wait time, non-payload activity, etc) goes into the "sequential-only" fraction.
When using several processors, one of them makes the sequential calculation, the others are waiting
(use the same amount of time). So, when calculating the speedup, one calculates

\begin{equation}
S=\frac{(1-\alpha)+\alpha}{(1-\alpha)+\alpha/k} =\frac{k}{k(1-\alpha)+\alpha}
\end{equation}
hence the  efficiency
\begin{equation}
\frac{S}{k}=\frac{1}{k(1-\alpha)+\alpha}
\end{equation}

This explains the behavior of diagram $\frac{S}{k}$ in function of $k$ in figures above:
the more processors, the lower efficiency.
In the case of Fig.~\ref{fig:KarpDiagram2},  ($1-\alpha$) is in the order of $10^{-3}$,
so the efficiency decreases to $0.5$ at $10^{3}$ processors, while
in the case of Fig.~\ref{fig:CompilerDiagram},  ($1-\alpha$) is $\approx 10^{-1}$,
so the efficiency decreases to $0.5$ at $10^{1}$ processors.
This is why Amdahl made  his very reasonable conclusion:"
\emph{the effort expended
	on achieving high parallel processing rates is wasted unless it is accompanied by achievements in
	sequential processing rates of very nearly the same magnitude}"~\cite{AmdahlSingleProcessor67}.

\section{Conclusions}
\label{sec:conclusion}
With the spread of both multi-core architectures,
using different parallelization solutions (like different networking or reconfigurable connection of cores, etc.)
and parallelizing the formerly sequential code
either with programmers' effort or using 
parallelizing compilers like the one by \cite{CompilerInfrastructure:2014}, it becomes more and more
important problem to characterize quantitatively the performance of the parallelization. 
Through inverting the formula known as Amdahl's law, 
and re-interpreting the comprised quantities,
such a figure of merit was derived. 
This experimental quantity correctly describes the performance
of parallelization, allowing to characterize the performance
of programmers or parallelizing compilers (see Fig.~\ref{fig:CompilerDiagram}),
different architectural solutions with many processors
 (see Fig.~\ref{fig:KarpDiagram1}),
different algorithms in function of the number of the processors
 (see Fig.~\ref{fig:KarpDiagram2}),
as well as describing
the performance of the network connection during running the task,
or quantifying the synchronization method used between the computing units. 
The introduced merit seems to be an adequate measure 
of the performance of the technology used for parallelization,
unlike the formerly used quantity (speedup divided by the number
of computing units).

\section*{References}


  \bibliographystyle{elsarticle-harv} 
  \bibliography{ParallelTOPC}


%
%
%
\end{document}